\documentclass[11pt,twoside]{article}
\usepackage{graphicx}
\usepackage{amsmath}
\usepackage{amssymb}
\usepackage{revsymb}
\usepackage{cite}
\usepackage{color}

\begin{document}
\newcommand{\pst}{\hspace*{1.5em}}

\newcommand{\rigmark}{\em Journal of Russian Laser Research}
\newcommand{\lemark}{\em Volume 30, Number 5, 2009}

\newcommand{\be}{\begin{equation}}
\newcommand{\ee}{\end{equation}}
\newcommand{\bm}{\boldmath}
\newcommand{\ds}{\displaystyle}
\newcommand{\bea}{\begin{eqnarray}}
\newcommand{\eea}{\end{eqnarray}}
\newcommand{\ba}{\begin{array}}
\newcommand{\ea}{\end{array}}
\newcommand{\arcsinh}{\mathop{\rm arcsinh}\nolimits}
\newcommand{\arctanh}{\mathop{\rm arctanh}\nolimits}
\newcommand{\bc}{\begin{center}}
\newcommand{\ec}{\end{center}}

\newcommand{\ket}[1]{|{#1}\rangle}
\newcommand{\bra}[1]{\langle{#1}|}
\newcommand{\ketbras}[3]{\ket{#1}_{#3}\hspace*{-0.2mm}\bra{#2}}
\renewcommand{\d}{\text{d}}
\newcommand{\e}{\text{e}}
\renewcommand{\i}{\text{i}}
\renewcommand{\Re}{\mathop{\text{Re}}\nolimits}
\renewcommand{\Im}{\mathop{\text{Im}}\nolimits}
\newcommand{\tr}{\mathop{\text{tr}}\nolimits}
\newcommand{\Tr}{\mathop{\text{Tr}}\nolimits}

\thispagestyle{plain}

\label{sh}


\begin{center}{\Large\bf
Thwarted dynamics by partial projective measurements
}\end{center}

\bigskip
\begin{center}{\bf
Bruno Bellomo$^{1*}$, Giuseppe Compagno$^1$, Hiromichi Nakazato$^2$ and Kazuya Yuasa$^3$
}\end{center}

\begin{center}
{\it
$^1$CNISM \& Dipartimento di Scienze Fisiche ed Astronomiche,
Universita di Palermo \\ via Archirafi 36, 90123 Palermo, Italy

\smallskip
$^2$Department of  Physics, Waseda University, Tokyo 169-8555, Japan

\smallskip
$^3$Waseda Institute for Advanced Study, Waseda University, Tokyo 169-8050, Japan
}

\smallskip
$^*$Corresponding author e-mail:~~~bruno.bellomo~@~gmail.com\\
\end{center}
\begin{abstract}\noindent
The dynamics of a system, made of a particle interacting with a field mode,
``thwarted'' by the action of repeated projective measurements on the
particle, is examined.
The effect of the partial measurements is discussed by comparing it with the dynamics in the absence of the measurements.
\end{abstract}


\section{Introduction}
\pst
For a generic bipartite system made of two interacting subsystems, ``partial'' measurements repeatedly performed on one of the two subsystems, projecting it onto a certain state, strongly affects the dynamics of the other.
Under certain conditions, the non-measured system is driven toward a pure state (distilled) irrespectively of its initial (in general, mixed) state \cite{1}.
This procedure would be useful in the context of quantum information for initializing quantum systems and would be utilized to prepare entangled states \cite{3,4}.
It is not so obvious however whether this scheme works also for systems with continuous spectra.
It has been shown, with a specific model, that it is actually possible to distill a state of a system by repeatedly performing measurements on another system with a \textit{continuous} spectrum, interacting with the former \cite{5}.

In this paper, we analyze this effect in more detail, by \textit{exactly} computing the dynamics of the target system under such repeated measurements.
The exact formulas also allow us to discuss the quantum Zeno effect and the quantum Zeno dynamics \cite{ref:QZE-review-PaoloSaverio} under the repeated partial measurements, without resort to approximations.

\section{Model and distillation}\label{par:Extraction of a squeezed field mode state}
\pst
The system we are going to study is the same as in Ref.\ \cite{5}.
It  consists of a particle $P$ of mass $m$ linearly interacting with a single field mode $F$ of frequency $\omega$.
The Hamiltonian reads
\begin{equation}
  \hat{H}=\frac{\hat{p}^2}{2m}+\hbar\omega
 \left( \hat{a}^{\dag}\hat{a}+\frac{1}{2}\right)+
   g \hat{p}(
  \hat{a}^{\dag}
  +\hat{a}
  ),
\end{equation}
where $\hat{p}$ is the momentum operator of the particle,
$\hat{a}$ and $\hat{a}^{\dag}$ are the
annihilation and creation operators of the field mode and $g$ is the
coupling constant between them.
At time $t=0$, the particle $P$ is prepared in a Gaussian state
\begin{equation}
\ket{\Phi_0}
=\int dp\,c_p\ket{p},\qquad
c_p=\frac{1}{\sqrt[4]{2\pi(\Delta p_0)^2}}\exp\!\left(
-\frac{p^2}{4(\Delta p_0)^2}
\right),
\label{eqn:Gauss}
\end{equation}
while the field mode $F$ is in an arbitrary (mixed) state $\hat{\rho}_F(0)$.
During the unitary dynamics of the total system $\hat{U}(\tau)=e^{-i\hat{H}\tau}$, the particle $P$ is repeatedly measured at intervals $\tau$ to confirm it to be in its initial state.
Each time, $P$ is projected onto the Gaussian state $|\Phi_0\rangle$.
This projection is partial, because only the particle $P$ is set back to its initial state, while the field $F$ is not.
Under the repeated actions of such measurements, the density matrix of the field mode $F$ just after the $N$th measurement reads, as a function of $N$,
\begin{equation}\label{contracted evolution operator e field density matrix evolution e probability}
     \hat{\rho}_F(N \tau)=\hat{V}^N\hat{\rho}_F(0)
      \hat{V}^{\dagger N}/P(N \tau), \qquad
P(N \tau)=\Tr_F\{\hat{V}^N\hat{\rho}_F(0)
  \hat{V}^{\dagger N} \}.
\end{equation}
Here
\begin{equation}
\hat{V}=\langle \Phi_{0}  | \hat{U}(\tau) |\Phi_{0}\rangle
\end{equation}
is the projected evolution operator between two consecutive measurements on $P$.
The normalization factor $P(N \tau)$ represents the probability of finding $P$ in its initial state in every measurement and thus gives the probability to obtain the state $\hat{\rho}_F(N \tau)$; it represents the survival probability of the initial Gaussian state $\ket{\Phi_0}$ of $P$ under the repeated measurements.
For the present setup, $\hat{U}(\tau)$ is explicitly cast into \cite{5}
\begin{equation}
   \hat{U}(\tau)=e^{-\frac{i}{\hbar}\frac{\hat{p}^2 \tau}{2 m}\,\bigl(1 -\frac{2 mg^2
 }{ \hbar\omega }
   \frac{ \omega \tau-
  \sin \omega \tau}{\omega \tau}\bigr)}	
  e^{-i \omega \tau\,\bigl( \hat{a}^{\dag}\hat{a}+\frac{1}{2}\bigr)}
 e^{\hat{p}(g_\tau\hat{a}^{\dag} -g_\tau^*\hat{a})},\qquad
g_\tau= g\frac{1-e^{i\omega \tau}}{\hbar \omega},
\label{eqn:U}
\end{equation}
and $\hat{V}$ is given by \cite{5}
\begin{gather}
\hat{V}
=M\exp\!\left\{
\frac{ \ln (q-\sqrt{q^2-1})}{\sqrt{q^2-1}}G\left[\frac{\hat{a}^{\dag2}+ \hat{a}^2}{2} +\tilde{q} \left(\hat{a}^\dag \hat{a}+\frac{1}{2}\right) \right] \right\},
\label{MFG}\\
\begin{cases}
\medskip
\displaystyle
M= \frac{1}{\sqrt{1+i\bar{\Delta}_p^2\bar{\tau}\left[1-2\bar{g}^2\left(1-\frac{\sin \bar{\tau}}{\bar{\tau}}\right)\right]}}, \\
\displaystyle
    G  = 2M^2\bar{g}^2\bar{\Delta}_p^2 (1-\cos \bar{\tau}),
\end{cases}\qquad
\begin{cases}
\medskip
\displaystyle
q=
\cos \bar{\tau}
+i G\sin\bar{\tau}, \\
    \displaystyle
\tilde{q}= \cos  \bar{\tau} +\frac{i}{G}\sin  \bar{\tau},
\end{cases}
\end{gather}
where $\bar{\tau}=\omega \tau$, $\bar{g}=g \sqrt{m/\hbar \omega}$, $\bar{\Delta}_p=\Delta p_0/\sqrt{m\hbar \omega}$ are the dimensionless parameters of the system.

At this point, it is possible to predict the final state of the field mode $F$ in the large $N$ limit keeping $\tau$ finite.
It is shown \cite{5} that the spectrum of $\hat{V}$ is discrete and its largest (in magnitude) eigenvalue is unique and nondegenerate \cite{1}, and the corresponding eigenstate is a squeezed state $\ket{\xi}$
with
\begin{equation}
\xi=  re^{i\varphi},\qquad
 r=\tanh^{-1}|\zeta|,\quad
 \varphi= \arg \zeta,\qquad
 \zeta=\tilde{q}\pm\sqrt{\tilde{q}^2-1},
\end{equation}
where the sign in $\zeta$ must be chosen so that $|q\pm\sqrt{q^2-1}|=|q\mp\sqrt{q^2-1}|^{-1}<1$ is satisfied \cite{5}.
Therefore, the field mode $F$ asymptotically approaches \cite{1,5}
\begin{equation}
\hat{\rho}_F(N \tau)\xrightarrow{\text{large\ } N} \ket{\xi}\bra{\xi}.
\end{equation}
The field mode $F$ is distilled into the pure squeezed state $\ket{\xi}$ from an arbitrary initial state $\hat{\rho}_F(0)$.

\section{Exact evolution under the repeated measurements}
\pst
Since the projected evolution operator $\hat{V}$ in (\ref{MFG}) is quadratic in the field-mode operators, it is even possible to derive the exact formula for the evolution of the field mode $F$ under the repeated measurements on the particle $P$, for any finite $N$.
It is convenient to work in the $P$-representation \cite{8}.
In general, the initial state of the field mode $F$ is expressed as
\begin{equation}
    \hat{\rho}_F(0)=\int d ^2 \alpha\,  \Pi(\alpha,\alpha^*) |\alpha\rangle
\langle \alpha|.
\end{equation}
It is easy to compute the evolution from a coherent state $\ket{\alpha}$ of the field mode $F$ under the repeated actions of $\hat{V}$.
First, the $N$th power of $\hat{V}$ given in (\ref{MFG}) is factorized as
\begin{equation}\label{VpowerN}
\hat{V}^N
=M^N e^{-\zeta_N\hat{a}^{\dag2}/2}e^{-\kappa_N(\hat{a}^\dag \hat{a}+1/2)}e^{-\zeta_N\hat{a}^2/2},
\end{equation}
\begin{equation}
\begin{cases}
\smallskip
\displaystyle
\zeta_N=\frac{G}{G \tilde{q}-\sqrt{q^2-1}\coth[N \ln (q-\sqrt{q^2-1})]},
\\
\displaystyle
\kappa_N=\ln\biggl(
\cosh[N \ln (q-\sqrt{q^2-1})] -\frac{G \tilde{q}}{\sqrt{q^2-1}} \sinh[N \ln  (q-\sqrt{q^2-1})]
\biggr).
\end{cases}
\end{equation}
Then, its action on the coherent state $\ket{\alpha}$ yields
\begin{equation}
\label{VN action on alfa}
   \hat{V}^N|\alpha\rangle
   =M_N\hat{D}(\alpha_N)\hat{S}(\xi_N)
   |0\rangle=
   M_N(\alpha)|\alpha_N,\xi_N\rangle ,
\end{equation}
\begin{equation}
\begin{cases}
\displaystyle
  \xi_N = r_Ne^{i \varphi_N},\\
\displaystyle
\alpha_N=( \alpha e^{-\kappa_N}-\alpha^*\zeta_Ne^{-\kappa_N^*})\cosh^2 r_N,
\end{cases}
\quad
\begin{cases}
\displaystyle
  r_N=\tanh^{-1}|\zeta_N|,\\
   \varphi_N=\arg \zeta_N,
   \end{cases}
\label{eqn:XiAlphaN}
\end{equation}
\begin{equation}
  M_N(\alpha) = M^N\sqrt{\cosh r_N} \exp\! \left(
 -\frac{1}{2}\,\Bigl[
 \kappa_N+|\alpha|^2(1-e^{-2\Re \kappa_N}\cosh^2 r_N)+\alpha^2(\zeta_N+\zeta_N^*e^{-2\kappa_N}\cosh^2 r_N)
\Bigr]\right),
 \end{equation}
 where $\hat{D}(\alpha)=e^{\alpha\hat{a}^\dag-\alpha^*\hat{a}}$ and $\hat{S}(\xi)=e^{-(\xi\hat{a}^{\dag2}+\xi^*\hat{a}^2)/2}$ are the displacement and squeezing operators, respectively, while $|\alpha_N,\xi_N\rangle$ is a squeezed coherent state \cite{8}.
Therefore, we immediately get
\begin{equation}\label{survival probability evoluzione campo expansion}
    \hat{\rho}_F(N \tau) =\frac{1}{P(N \tau)}\int d ^2 \alpha\,  |M_N(\alpha) |^2 \Pi(\alpha,\alpha^*)
      |\alpha_N,\xi_N\rangle \langle\alpha_N,\xi_N|  ,\quad
    P(N \tau)=\int d ^2 \alpha\,  |M_N (\alpha)|^2 \Pi(\alpha,\alpha^*)
   .
\end{equation}
These are exact for any $N$.
Various quantities of interest can be computed on the basis of these formulas.
For instance, the average number of field quanta $  \langle \hat{a}^{\dag}\hat{a}
    \rangle_{N\tau}=\Tr_F\{\hat{\rho}_F(N \tau)\hat{a}^\dag\hat{a}\}$ and the fidelity $F(N \tau)=\langle \xi| \hat{\rho}_F(N \tau)|\xi \rangle  $  of $\hat{\rho}_F(N \tau)$ with respect to the target squeezed state $\ket{\xi}$ are  given, respectively, by
\begin{gather}\label{numero medio di fotoni and fidelity expansion}
\langle \hat{a}^{\dag}\hat{a}
    \rangle_{N\tau} = \sinh^2r_N+\frac{1}{P(N \tau)}\int d ^2 \alpha\,|\alpha_N|^2  |M_N(\alpha) |^2
    \Pi(\alpha,\alpha^*)
    ,\\
F(N \tau) =\frac{1}{P(N \tau)}\int d ^2 \alpha\,  |M_N(\alpha) |^2 \Pi(\alpha,\alpha^*) |\langle \xi|
      \alpha_N,\xi_N \rangle |^2 .
\end{gather}

\section{Evolution in the absence of the measurements}
\label{par:Comparison with the no measurement case}
\pst
Before proceeding to analyze the evolution of the system under the repeated measurements in detail, let us collect the corresponding formulas for the evolution of the system in the absence of the measurements.
For instance, the survival probability of the initial Gaussian state $\ket{\Phi_0}$ of the particle $P$ at time $t$ in the absence of the measurements,  $P^{(0)}(t)=\Tr_F\{\bra{\Phi_0}\hat{U}(t)[\ket{\Phi_0}\bra{\Phi_0}\otimes\hat{\rho}_F(0)]\hat{U}^\dag(\tau)\ket{\Phi_0}\}$, is obtained by substituting $N\to1$ and $\tau\to t$ in Eq.\ (\ref{survival probability evoluzione campo expansion}) as
\begin{eqnarray}\label{p without measurements}
   P^{(0)}(t)=\int d ^2 \alpha\,\frac{|M|^{2}}{\sqrt{1+2\Re G}}\exp\! \left[ -\frac{\Re G}{1+2\Re G}\left(\alpha e^{-i\omega t/2}+\alpha^* e^{i\omega t/2}\right)^2\right]
    \Pi(\alpha,\alpha^*) .
\end{eqnarray}
As for the reduced density matrix of the field mode $F$ in the absence of the measurements, $\hat{\rho}_F^{(0)}(t)=\Tr_P\{\hat{U}(t)[\ket{\Phi_0}\bra{\Phi_0}\otimes\hat{\rho}_F(0)]\hat{U}^\dag(\tau)\}$, it is obtained by directly computing the trace over the particle's degrees of freedom with the exact expression of the evolution operator of the total system, $\hat{U}(t)$ given in (\ref{eqn:U}), as
\begin{equation}\label{elementi matrice ridotta campo2bis}
   \hat{\rho}_F^{(0)}(t)= \int d ^2 \alpha\,\Pi(\alpha,\alpha^*)    \int d
  p\,
  |c_p|^2 |(\alpha+p g_t) e^{-i \omega t}\rangle
   \langle(\alpha+p g_t) e^{-i \omega t}|,
\end{equation}
where $c_p$ and $g_t$ are given in (\ref{eqn:Gauss}) and (\ref{eqn:U}), respectively.
Then, the average number of field quanta $\langle \hat{a}^{\dag}\hat{a}\rangle_t^{(0)}=\Tr_F\{\hat{\rho}_F^{(0)}(t) \hat{a}^{\dag}\hat{a}\}$ is estimated to be
\begin{equation}\label{media fotonicalcolo5}
  \langle\hat{a}^{\dag}\hat{a}\rangle_t^{(0)} =  \int d ^2 \alpha\,  \Pi(\alpha,\alpha^*) \,\Bigl(
|\alpha|^2 +(\Delta p_0)^2|g_t|^2
\Bigr).
\end{equation}

\section{Initial coherent state}
\pst
Let us now look at the evolution of the system under the repeated measurements on $P$ in detail, for the initial coherent state $\ket{\alpha_0}$ of the field mode $F$.
In this case, by putting $\Pi(\alpha,\alpha^*)=\delta^2(\alpha-\alpha_0)$, the extracted state of the field mode $F$ and the survival probability of the initial Gaussian state of $P$, both  given in (\ref{survival probability evoluzione campo expansion}), are reduced to
\begin{equation}
    \hat{\rho}_F(N \tau) =|\alpha_N,\xi_N\rangle \langle\alpha_N,\xi_N|  ,\qquad
    P(N \tau)=|M_N (\alpha_0)|^2.
\label{eqn:RhoPcoherent}
\end{equation}
After each measurement, the field mode is in a squeezed  coherent state, with the squeezing parameter and the coherent-state amplitude evolve according to Eq.\ (\ref{eqn:XiAlphaN}).
The average number of field quanta reads
\begin{eqnarray}\label{numero medio di fotoni}
   \langle \hat{a}^{\dag}\hat{a}
    \rangle_{N\tau} =\sinh^2  r_N+|\alpha_N|^2 =\frac{|\zeta_N|}{1-|\zeta_N|^2}
    +|\alpha_N|^2
     .
\end{eqnarray}

We consider two different limits: (i) distillation limit $N\to\infty$ keeping $\tau$ finite and (ii) Zeno limit $N\to\infty$ keeping $t=N\tau$ fixed.
In the former case (i), the state of the field mode $F$ is led to the squeezed state $\hat{\rho}_F(N \tau)\to\ket{\xi}\bra{\xi}$, since the formulas in (\ref{eqn:XiAlphaN}) suggest $\alpha_N\to0$ and $\xi_N\to\xi$.
The formulas in (\ref{eqn:RhoPcoherent}), valid for any finite $N$, tell us how the state of $F$ approaches the final state $\ket{\xi}$.
In particular, the fidelity to the target squeezed state $\ket{\xi}$ is estimated to be
\begin{eqnarray}\label{fidelity starting from alfa}
F(N \tau)    = |\langle\xi|\alpha_N,\xi_N\rangle|^2=
   \frac{ \exp\! \left[
    -|\alpha_N|^2 \frac{
    \cosh 2 r +\cosh2r_N  + \sinh2 r
    \cos(\varphi-2\theta_{N})+ \sinh 2 r_N \cos
    (\varphi_{N}-2\theta_{N})}{1+\cosh 2 r \cosh 2 r_N-
    \sinh2r \sinh 2 r_N\cos (\varphi-\varphi_{N}) } \right]
    }{\sqrt{1+\cosh (2 r) \cosh (2 r_N)-
    \sinh (2r) \sinh (2 r_N)\cos (\varphi-\varphi_{N})}/\sqrt{2}},
\end{eqnarray}
where $\theta_N=\arg\alpha_N$.
It approaches unity $F(N \tau)\to1$, while the probability decays to zero  $P(N \tau)\to0$ as the number of the measurements $N$ increases.
These exact formulas support the argument in Sec.\ 2 and Ref.\ \cite{5}.
The average number of field quanta accordingly reaches the average number in the squeezed state $\ket{\xi}$, i.e., $\langle
\hat{a}^{\dag}\hat{a} \rangle_{N\tau} \rightarrow \sinh^2 r$.

\begin{figure}[b]
\bc
\includegraphics[width=5.7cm]{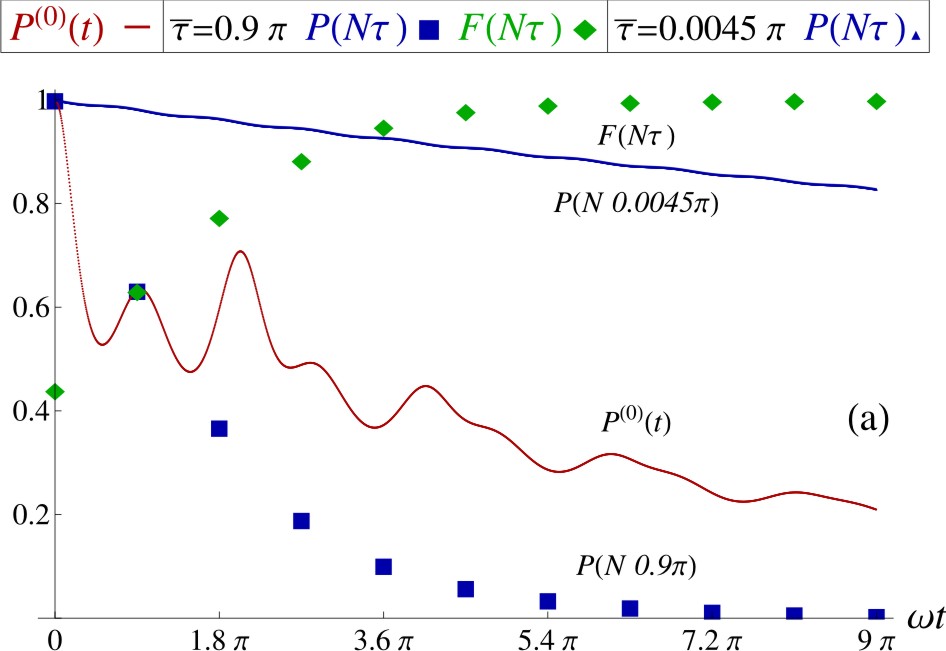}
\includegraphics[width=5.7cm]{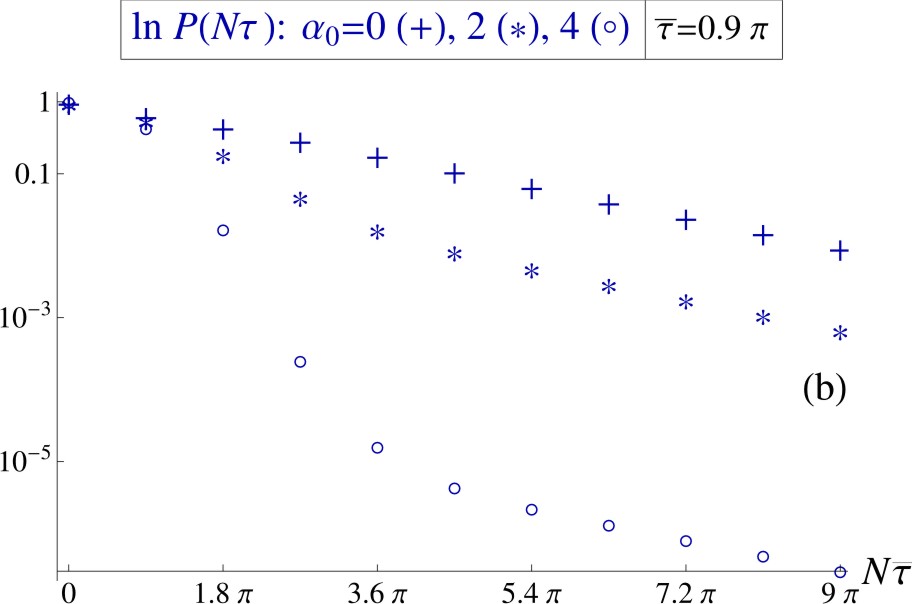}
\includegraphics[width=5.7cm]{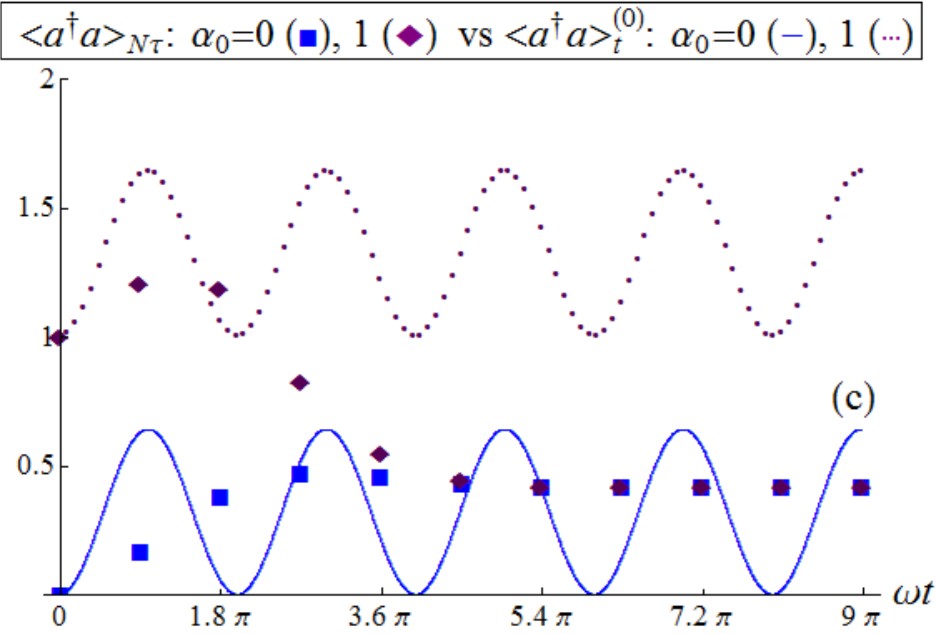}
\ec
\vspace{-4mm}
\caption{\label{figcoherent}
(a) Survival probability $P(N \tau)$ under the repeated measurements for $\bar{\tau}=0.9\pi$ (squares) and $0.0045\pi$ (points merged to solid line), in comparison to $P^{(0)}(t)$ in the absence of the measurements (solid), together with the evolution of the fidelity $F(N \tau)$ (diamonds) of the state of the field model $F$ with respect to the squeezed state $\ket{\xi}$ under the repeated measurements. The amplitude of the initial coherent state is set $\alpha_0=1$.
(b) $\ln P (N \tau)$ is shown for $\alpha_0=0,2,4$ (from top to bottom) with $\bar{\tau}=0.9\pi$ fixed.
(c) Average number of field quanta $\langle\hat{a}^\dag\hat{a}\rangle_{N\tau}$ under the repeated measurements  for $\alpha_0=0$ (squares) and $1$ (diamonds) with $\bar{\tau}=0.9\pi$ fixed, in comparison  with  $\langle\hat{a}^\dag\hat{a}\rangle_t^{(0)}$ in the absence of the measurements for $\alpha_0=0$ (solid) and $\alpha_0=1$ (dotted).
The other parameters are $\bar{g}=1$ and $\bar{\Delta}_p=0.4$ in  all panels.
}
\end{figure}

In Fig.\ \ref{figcoherent}(a), the fidelity $F(N \tau)$  (diamonds) and the probability $P(N \tau)$ under the repeated measurements  (squares), as well as the probability $P^{(0)}(t)$ in the absence of the measurements (solid), are plotted for the initial coherent state $\ket{\alpha_0}$ of the field mode $F$, with parameters $\bar{g}=1$, $\bar{\Delta}_p=0.4$, $\alpha_0=1$ and $\bar{\tau}=0.9 \pi$.
The fidelity $F(N \tau)$ (diamonds) quickly increases after a small number of measurements, before the probability $P(N \tau)$ (squares) decays out completely.
In Fig.\ \ref{figcoherent}(c), the evolution of the average number of field quanta $\langle
\hat{a}^{\dag}\hat{a} \rangle_{N\tau}$ is shown  for $\alpha_0=0$ (squares) and $\alpha_0=1$ (diamonds).
It  approaches the constant $\sinh^2r$, while it oscillates as $\langle\hat{a}^{\dag}\hat{a}\rangle_t^{(0)}=  |\alpha_0|^2+4\bar{g}^2\bar{\Delta}_p^2\sin^2(\omega t/2)  $, between $|\alpha_0|^2$ and $ |\alpha_0|^2+ 4\bar{g}^2 \bar{\Delta}_p^2$, in the absence of the measurements, as shown for $\alpha_0=0$ (solid) and $\alpha_0=1$ (dotted), during which the particle $P$ exchanges energy with the field mode $F$.

In the Zeno limit (ii), on the other hand, the explicit formula in (\ref{eqn:RhoPcoherent}) shows that the survival probability becomes $P(N \tau)\to1$.
That is, the particle $P$ is frozen in the initial Gaussian state $\ket{\Phi_0}$ and never evolves (the quantum Zeno effect \cite{ref:QZE-review-PaoloSaverio}).
However, the field mode $F$ can evolve in the ``Zeno subspace'' specified by the projection operator $\hat{P}_Z=\ket{\Phi_0}\bra{\Phi_0}\otimes\openone_F$ \cite{ref:QZE-review-PaoloSaverio,ref:ZenoDynamics}.
The exact formula (\ref{eqn:RhoPcoherent}) actually shows that in the Zeno limit the field mode $F$ evolves as $\hat{\rho}_F(t) =|\alpha_0e^{-i\omega t}\rangle \langle\alpha_0 e^{-i\omega t}|$.
This is the ``quantum Zeno dynamics'' of the field mode $F$ in the Zeno subspace, under the frequent measurements on the particle $P$, and is exactly the dynamics driven by the ``Zeno Hamiltonian'' defined by  $\hat{H}_Z=\hat{P}_Z\hat{H}\hat{P}_Z=\ket{\Phi_0}\bra{\Phi_0}\otimes\hbar\omega\hat{a}^\dag\hat{a}$, supporting the generic argument in Refs.\ \cite{ref:QZE-review-PaoloSaverio,ref:ZenoDynamics}.

In Fig.\ \ref{figcoherent}(a) the survival probabilities $P(N \tau)$ with two different intervals between consecutive measurements, $\bar{\tau}=0.9 \pi$ (squares) and $0.0045 \pi$ (points merged to solid line), are shown, in comparison with the survival probability $P^{(0)}(t)$ in the absence of the measurements (solid).
The decay of the survival probability $P(N \tau)$  with a shorter interval $\tau$ is suppressed as manifestation of the quantum Zeno effect.
On the other hand, with a rather longer interval, the decay is accelerated  as compared to the natural decay in the absence of the measurements, which is called ``anti-Zeno effect'' \cite{ref:Kurizki} or ``inverse Zeno effect'' \cite{9}.

What is interesting in the present analysis is that the measurements frequently performed on the system are ``partial''.
Such explicit expressions with partial measurements  as (\ref{eqn:RhoPcoherent}) are usually absent in the literature, and the arguments often rely on delicate limiting procedures.
In addition, when a measurement projects a total system onto its pure state by a projection operator $\hat{\mathcal{O}}=\ket{\Psi}\bra{\Psi}$, the repeated applications of such measurements lead  to an exponential decay of the survival probability as $P(N \tau)=|\bra{\Psi}\hat{U}(\tau)\ket{\Psi}|^{2N}=[P^{(0)}(\tau)]^N$.
In the present case with the partial measurements, this is not the case: $\ln P(N \tau)$ does not exhibit a straight line.
This is shown in Fig.\ 1(b), where the curvature becomes more prominent with a larger $\alpha_0$ (from up to down).

\section{Concluding remarks}
\pst
In this paper, we have examined the distillation process studied in Ref.\ \cite{5} with a specific model, on the basis of the explicit and exact formulas that describe the evolution of the system under the repeated measurements.
The formulas also allow us to discuss the quantum Zeno effect and the quantum Zeno dynamics without resort to approximate treatment.
In the present study, we considered a single field mode  with a discrete spectrum, but the analysis could be extended to a field  with a continuous spectrum.
Many of the previous works on the quantum Zeno effect concern measurements which project the total system onto pure states.
Partial measurements as discussed here lead to nonexponential decays of the survival probabilities under repeated measurements, and such explicit and exact calculation would provide us with a new insight on the transition from Zeno to the inverse Zeno effect.

\end{document}